\documentclass[review]{elsarticle}

\usepackage{lineno,hyperref}
\modulolinenumbers[5]

\makeatletter
\def\mdseries@tt{m}
\makeatother
\usepackage[cache=false, draft]{minted}

\usepackage{graphicx}
\usepackage{float}
\usepackage{subcaption}
\usepackage{booktabs}
\usepackage{siunitx}
\usepackage{listings}
\usepackage{multirow}
\usepackage{amsmath}
\usepackage{xcolor}

\definecolor{my_white}{rgb}{1, 1, 1}                
\definecolor{my_black}{rgb}{0, 0, 0}                
\definecolor{my_blue}{rgb}{0, 0.125, 0.376}         
\definecolor{my_green}{rgb}{0.12, 0.3, 0.17}        
\definecolor{my_violet}{rgb}{0.44, 0.16, 0.39}      

\lstdefinestyle{Bash}{
  backgroundcolor=\color{my_white},   
  basicstyle=\footnotesize,        
  breakatwhitespace=false,         
  breaklines=true,                 
  captionpos=b,                    
  commentstyle=\color{my_green},   
  frame=single,	                   
  keepspaces=true,                 
  keywordstyle=\color{my_blue},    
  language=bash,                   
  numbers=none,                    
  numbersep=5pt,                   
  numberstyle=\tiny\color{my_black},   
  rulecolor=\color{my_black},      
  showspaces=false,                
  showstringspaces=false,          
  showtabs=false,                  
  stepnumber=1,                    
  stringstyle=\color{my_violet},   
  tabsize=2,	                   
  morekeywords={git, sudo, apt, get, pip2, okular, wxparaver},             
  title=\lstname                   
}

\lstdefinestyle{PyCOMPSs}{
  backgroundcolor=\color{my_white},   
  basicstyle=\footnotesize,        
  breakatwhitespace=false,         
  breaklines=true,                 
  captionpos=None,                    
  commentstyle=\color{my_green},   
  frame=single,	                   
  keepspaces=true,                 
  keywordstyle=\color{my_blue},    
  language=Python,                 
  numbers=none,                    
  numbersep=5pt,                   
  numberstyle=\tiny\color{my_black},   
  rulecolor=\color{my_black},      
  showspaces=false,                
  showstringspaces=false,          
  showtabs=false,                  
  stepnumber=1,                    
  stringstyle=\color{my_violet},   
  tabsize=2,	                   
  morekeywords={@task, @constraint, @parallel, compss_wait_on},             
}

\title{A Programming Model for Hybrid Workflows: combining Task-based Workflows and Dataflows all-in-one}

\author[bsc]{Cristian Ramon-Cortes}
\author[bsc]{Francesc Lordan}
\author[bsc]{Jorge Ejarque}
\author[bsc]{Rosa M. Badia}

\address[bsc]{Barcelona Supercomputing Center (BSC)}

\begin{document}

    This manuscript has been accepted at the Future Generation Computer Systems (FGCS). \\
    DOI: 10.1016/j.future.2020.07.007 \\
    This manuscript is licensed under CC-BY-NC-ND. \\
    
    \begin{abstract}

In the past years, e-Science applications have evolved from large-scale simulations executed in a single cluster to more complex workflows where these simulations are combined with High-Performance Data Analytics (HPDA). To implement these workflows, developers are currently using different patterns; mainly task-based and dataflow. However, since these patterns are usually managed by separated frameworks, the implementation of these applications requires to combine them; considerably increasing the effort for learning, deploying, and integrating applications in the different frameworks. 

This paper tries to reduce this effort by proposing a way to extend task-based management systems to support continuous input and output data to enable the combination of task-based workflows and dataflows (Hybrid Workflows from now on) using a single programming model. Hence, developers can build complex Data Science workflows with different approaches depending on the requirements. To illustrate the capabilities of Hybrid Workflows, we have built a Distributed Stream Library and a fully functional prototype extending COMPSs, a mature, general-purpose, task-based, parallel programming model. The library can be easily integrated with existing task-based frameworks to provide support for dataflows. Also, it provides a homogeneous, generic, and simple representation of object and file streams in both Java and Python; enabling complex workflows to handle any data type without dealing directly with the streaming back-end.

During the evaluation, we introduce four use cases to illustrate the new capabilities of Hybrid Workflows; measuring the performance benefits when processing data continuously as it is generated, when removing synchronisation points, when processing external real-time data, and when combining task-based workflows and dataflows at different levels. The users identifying these patterns in their workflows may use the presented uses cases (and their performance improvements) as a reference to update their code and benefit of the capabilities of Hybrid Workflows. Furthermore, we analyse the scalability in terms of the number of writers and readers and measure the task analysis, task scheduling, and task execution times when using objects or streams. 

\end{abstract}

    
    \maketitle

\section{Introduction}
\label{sec:introduction}

%
%
%

For many years, large-scale simulations, High-Performance Data Analytics (HPDA), and simulation workflows have become a must to progress in many scientific areas such as life, health, and earth sciences. In such a context, there is a need to adapt the High-Performance infrastructure and frameworks to support the needs and challenges of workflows combining these technologies~\cite{web:intersection_ai_hpc_hpda}. 

Traditionally, developers have tackled the parallelisation and distributed execution of these applications following two different strategies. On the one hand, task-based workflows orchestrate the execution of several pieces of code (\textit{tasks}) that process and generate data values. These tasks have no state and, during its execution, they are isolated from other tasks; thus, task-based workflows consist of defining the data dependencies among tasks. On the other hand, dataflows assume that tasks are persistent executions with a state that continuously receive/produce data values (streams). Through dataflows, developers describe how the tasks communicate to each other. 

\begin{figure}[!htb]
  \centering
  \includegraphics[width=0.75\linewidth]{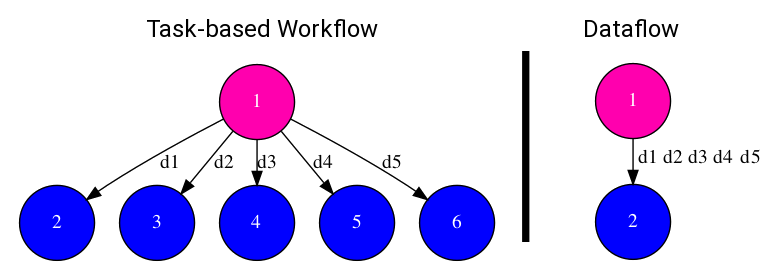}
  \caption{Representation of the same workflow using task-based and dataflow patterns (left and right respectively).}
  \label{fig:tf_df_graphs}
\end{figure}

Regardless of the workflow type, directed graphs are a useful visualisation and management tool. Figure~\ref{fig:tf_df_graphs} shows the graph representation of a task-based workflow (left) and its equivalent dataflow (right). The task dependency graph consists of a producer task (coloured in pink) and five consumer tasks (coloured in blue) that can run in parallel after the producer completes. The dataflow graph also has a producer task (coloured in pink), but one single stateful consumer task (coloured in blue) which processes all the input data sequentially (unless the developer internally parallelises it). Rather than waiting for the completion of the producer task to process all its outputs, the consumer task can process the data as it is generated.

The first contribution of this paper is the proposal of a single hybrid programming model capable of executing task-based workflows and dataflows simultaneously.
For that purpose, we extend task-based frameworks to support continuous input and output data, and enable the combination of task-based workflows and dataflows (Hybrid Workflows from now on) using the same programming model.
Moreover, it allows developers to build complex Data Science pipelines with different approaches depending on the requirements. The evaluation (Section~\ref{sec:evaluation}) demonstrates that the use of Hybrid Workflows has significant performance benefits when identifying some patterns in task-based workflows; e.g., when processing data continuously as it is generated, when removing synchronisation points, when processing external real-time data, and when combining task-based workflows and dataflows at different levels. Also, notice that using a single programming model frees the developers from the burden of deploying, using, and communicating different frameworks inside the same workflow.

The second contribution presented in this paper is a Distributed Streaming Library that can be easily integrated with existing task-based frameworks to provide support for dataflows. The library provides a homogeneous, generic, and simple representation of a stream for enabling complex workflows to handle any kind of data without dealing directly with the streaming back-end. At its current state, the library supports file streams through a custom implementation, and object streams through Kafka~\cite{kafka}.

To validate and evaluate our proposal, we have built a prototype that helps us to illustrate the additional capabilities of Hybrid Workflows for complex Data Science applications. This prototype extends COMPSs~\cite{compss_softwareX, web:compss}, a mature, general-purpose programming model, and integrates the Distributed Streaming Library.

The rest of the paper is organised as follows. 
Section~\ref{sec:related_work} presents an overview of the related work, and Section~\ref{sec:background} introduces the baseline technology on which our solution is built. 
Next, Section~\ref{sec:architecture} describes the architecture of the Distributed Stream Library and its integration with COMPSs.
Section~\ref{sec:use_cases} details four use cases to illustrate the new available features, and
Section~\ref{sec:evaluation} evaluates our proposal measuring performance improvements at application level and the additional runtime overhead.
Finally, Section~\ref{sec:conclusion} concludes the paper and gives some guidelines for future work.
    
\section{Related Work}
\label{sec:related_work}

%
%
%
%

Nowadays, state-of-the-art frameworks typically focus on the execution of either task-based workflows or dataflows. Thus, next subsections provide a general overview of the most relevant frameworks for both task-based workflows and dataflows. Furthermore, since our prototype combines both approaches into a single programming model and allows developers to build Hybrid Workflows without deploying and managing two different frameworks, the last subsection details other solutions and compares them with our proposal.

\subsection{Task-based frameworks}

Although all the frameworks handle the tasks and data transfers transparently, there are two main approaches to define task-based workflows. On the one hand, many frameworks force developers to explicitly define the application workflow through a recipe file or a graphical interface.
FireWorks~\cite{fireworks, web:fireworks} defines complex workflows using recipe files in Python, JSON, or YAML. It focuses on high-throughput applications, such as computational chemistry and materials science calculations, and provides support arbitrary computing resources (including queue systems), monitoring through a built-in web interface, failure detection, and dynamic workflow management. 
Taverna~\cite{taverna, web:taverna} is a suite of tools to design, monitor, and execute scientific workflows. It provides a graphical user interface for the composition of workflows that are written in a Simple Conceptual Unified Flow Language (Scufl) and executed remotely by the Taverna Server to any underlying infrastructure (such as supercomputers, Grids or cloud environments). 
Similarly, Kepler~\cite{kepler, web:kepler} also provides a graphical user interface to compose workflows by selecting and connecting analytical components and data sources. Furthermore, workflows can be easily stored, reused, and shared across the community. Internally, Kepler's architecture is actor-oriented to allow different execution models into the same workflow. 
Also, Galaxy~\cite{galaxy, web:galaxy} is a web-based platform for data analysis focused on accessibility and reproducibility of workflows across the scientific community. The users define workflows through the web portal and submit their executions to a Galaxy server containing a full repertoire of tools and reference data. 
In an attempt to increase the interoperability between the different systems and to avoid the duplication of development efforts, Tavaxy~\cite{tavaxy} integrates Taverna and Galaxy workflows in a single environment; defining an extensible set of re-usable workflow patterns and supporting cloud capabilities. Although Tavaxy allows the composition of workflows using Taverna and Galaxy sub-workflows, the resulting workflow does not support streams nor any dataflow pattern. 

On the other hand, other frameworks implicitly build the task dependency graph from the user code. Some opt for defining a new scripting language to manage the workflow. These solutions force the users to learn a new language but make a clear differentiation between the workflow's management (the script) and the processes or programs to be executed. 
Nextflow~\cite{nextflow, web:nextflow} enables scalable and reproducible workflows using software containers. It provides a fluent DSL to implement and deploy workflows but allows the adaptation of pipelines written in the most common scripting languages.
Swift~\cite{swift, web:swift} is a parallel scripting language developed in Java and designed to express and coordinate parallel invocations of application programs on distributed and parallel computing platforms. Users only define the main application and the input and output parameters of each program, so that Swift can execute the application in any distributed infrastructure by automatically building the data dependencies. 

Other frameworks opt for defining some annotations on top of an already existing language. These solutions avoid the users from learning a new language but merge the workflow annotations and its execution in the same files. 
Parsl~\cite{web:parsl} evolves from Swift and provides an intuitive way to build implicit workflows by annotating "apps" in Python codes. In Parsl, the developers annotate Python functions (apps) and Parsl constructs a dynamic, parallel execution graph derived from the implicit linkage between apps based on shared input/output data objects. Parsl then executes apps when dependencies are met. Parsl is resource-independent, that is, the same Parsl script can be executed on a laptop, cluster, cloud, or supercomputer.
Dask~\cite{dask} is a library for parallel computing in Python. Dask follows a task-based approach being able to take into account the data-dependencies between the tasks and exploiting the inherent concurrency. Dask has been designed for computation and interactive data science and integration with Jupyter notebooks. It is built on the dataframe data-structure that offers interfaces to NumPy, Pandas, and Python iterators. Dask supports implicit, simple, task-graphs previously defined by the system (Dask Array or Dask Bag) and, for more complex graphs, the programmer can rely in the {\tt delayed} annotation that supports the asynchronous executions of tasks by building the corresponding task-graph. 
COMPSs~\cite{compss_softwareX, compss_servicess, web:compss} is a task-based programming model for the development of workflows/applications to be executed in distributed programming platforms. The task-dependency graph (or workflow) is generated at execution time and depends on the input data and the dynamic execution of the application. Thus, compared with other workflow systems that are based on the static drawing of the workflow, COMPSs offers a tool for building dynamic workflows, with all the flexibility and expressivity of the programming language.

\subsection{Dataflow frameworks}

Stream processing has become an increasingly prevalent solution to process social media and sensor devices data. On the one hand, many frameworks have been created explicitly to face this problem.
Apache Flink~\cite{web:flink} is streaming dataflow engine to perform stateful computations over data streams (i.e., event-driven applications, streaming pipelines or stream analytics). It provides exactly-once processing, high throughput, automated memory management, and advanced streaming capabilities (such as windowing). Flink users build dataflows that start with one or more input streams (sources), perform arbitrary transformations, and end in one or more outputs (sinks). 
Apache Samza~\cite{web:samza} allows building stateful applications for event processing or real-time analytics. Its differential point is to offer built-in support to process and transform data from many sources, including Apache Kafka, AWS Kinesis, Azure EventHubs, ElasticSearch, and HDFS. Samza users define a stream application that processes messages from a set of input streams, transforms them by chaining multiple operators, and emits the results to output streams or stores. Also, Samza supports at-least-once processing, guaranteeing no data-loss even in case of failures. 
Apache Storm~\cite{storm} a is distributed real-time computation system based on the master-worker architecture and used in real-time analytics, online machine learning, continuous computation, and distributed RPC, between others. Storm users define topologies that consume streams of data (spouts) and process those streams in arbitrarily complex ways (bolts), re-partitioning the streams between each stage of the computation however needed. Although Storm natively provides at-least-once processing, it also supports exactly-once processing via its high-level API called Trident.
Twitter Heron~\cite{heron} is a real-time, fault-tolerant stream processing engine. Heron was built as the Storm's successor, meaning that the topology concepts (spouts and bolts) are the same and has a compatible API with Storm. However, Heron provides better resource isolation, new scheduler features (such as on-demand resources), better throughput, and lower latency.

\subsection{Hybrid frameworks}

Apache Spark~\cite{spark} is a general framework for big data processing that was originally designed to overcome the limitations of MapReduce~\cite{map_reduce}. Among the many built-in modules, Spark Streaming~\cite{spark_streaming} is an extension of the Spark core to evolve from batch processing to continuous processing by emulating streaming via micro-batching. It ingests input data streams from many sources (e.g., Kafka, Flume, Kinesis, ZeroMQ) and divides them into batches that are then processed by the Spark engine; allowing to combine streaming with batch queries. Internally, the continuous stream of data is represented as a sequence of RDDs in a high-level abstraction called Discretized Stream (DStream). 

Notice that Spark is based on high-level operators (operators on RDDs) that are internally represented as a DAG; limiting the patterns of the applications. In contrast, our approach is based on sequential programming, which allows the developer to build any kind of application. Furthermore, micro-batching requires a predefined threshold or frequency before any processing occurs; which can be "real-time" enough for many applications, but may lead to failures when micro-batching is simply not fast enough. In contrast, our solution uses a dedicated streaming engine to handle dataflows; relying on streaming technologies rather than micro-batching and ensuring that the data is processed as soon as it is available. 

On the other hand, other solutions combine existing frameworks to support Hybrid Workflows. Asterism~\cite{asterism} is a hybrid framework combining \verb|dispel4py| and Pegasus at different levels to run data-intensive stream-based applications across platforms on heterogeneous systems. The main idea is to represent the different parts of a complex application as \verb|dispel4py| workflows which are, then, orchestrated by Pegasus as tasks. While the stream-based execution is managed by \verb|dispel4py|, the data movement between the different execution platforms and the workflow engine (submit host) is managed by Pegasus. Notice that Asterism can only handle dataflows inside task-based workflows (\verb|dispel4py| workflows represented as Pegasus' tasks), while our proposal is capable of orchestrating nested task-flows, nested dataflows, dataflows inside task-based workflows, and task-based workflows inside dataflows.

\section{Background}
\label{sec:background}

%
%
%

To the end of enabling the construction of Hybrid Workflows using the same programming model, we decided to extend an already existing task-based programming model and enable the support for dataflows. Since we have chosen COMPSs as the base workflow manager for our prototype, this section introduces the essential concepts for understanding its programming model and supporting runtime. The second part of this section briefly introduces Kafka since the default backend of the Distributed Streaming Library uses it to support object streams.

%
%
\subsection{COMPSs}
\label{subsec:compss}

COMP Superscalar (COMPSs) is a programming model based on sequential programming and designed to abstract developers away from the parallelisation and distribution details such as thread creation, synchronisation, data distribution, message passing or fault-tolerance. COMPSs is a task-based model; application developers select a set of methods whose invocations are considered tasks that will run asynchronously in distributed nodes. 

\begin{figure}[!htb]
  \centering
  \includegraphics[width=0.7\linewidth]{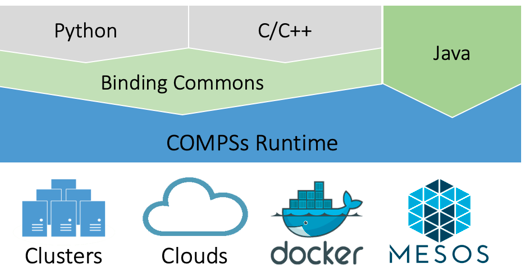}
  \caption{COMPSs overview.}
  \label{fig:compss_overview}
\end{figure}

As shown in Figure~\ref{fig:compss_overview}, Java is the native programming language to develop COMPSs applications; however, it also provides bindings for Python (PyCOMPSs~\cite{pycompss}) and C/C++~\cite{compsstango}. Its programming model is based on annotations that are used to choose class and object methods as tasks. These annotations can be split into two groups:

\begin{itemize}
  \item \textbf{Method Annotations:} Annotations added to the sequential code methods to define them as tasks and potentially execute them in parallel.
  \item \textbf{Parameter Annotations:} Annotations added to the parameters of an annotated method to indicate the direction (IN, OUT, INOUT) of the data used by a task.
\end{itemize}

The runtime~\cite{compss_servicess} system supporting the model follows a master-worker architecture. The master node, on which the main code of the application runs, orchestrates the execution of the applications and its tasks on the underlying infrastructure -- described in an XML configuration file. For that purpose, it intercepts calls to methods annotated as tasks and, for each detected call, it analyses the data dependencies with previous tasks according to defined parameter annotations. As the result of this analysis, the runtime builds a Directed Acyclic Graph (DAG) where nodes represent tasks and edges represent data dependencies between them; thus, the runtime infers the application parallelism. The master node transparently schedules and submits a task execution on a worker node handling the required data transfers. If a partial failure raises during a task execution, the master node handles it with job re-submission and re-schedule techniques.

COMPSs guarantees the portability across different computing platforms such as clusters, supercomputing machines, clouds, or container-managed infrastructures without modifying the application code. Also, the runtime allows the usage of third-party plugins by implementing a simple interface to extend its usage to new infrastructures or change the scheduling policy.

Furthermore, the COMPSs framework also provides a live-monitoring tool through a built-in web interface. For further details on executions, users can enable the instrumentation of their application using Extrae~\cite{web:extrae} and generate post-mortem traces that can be analysed with Paraver~\cite{web:paraver}.

\subsubsection{Java Syntax}

A COMPSs application in Java is composed of three parts:

\begin{itemize}
  \item \textbf{Main application:} Sequential code defining the workflow of the application. It must contain calls to class or object methods annotated as tasks so that, at execution time, they can be asynchronously executed in the available resources. 
  \item \textbf{Remote Methods:} Code containing the implementation of the tasks. This code can be in the same file than the application's main code or in one or more separate files.
  \item \textbf{Annotated Interface:} List of annotated methods that can be remotely executed as tasks. It contains one entry per task defining the \textit{Method Annotation}, the object or class method name, and one \textit{Parameter Annotation} per method parameter. 
\end{itemize}

Notice that COMPSs Java applications do not require any special API, pragma or construct since COMPSs instruments the application's code at execution time to detect the tasks defined in the annotated interface. Hence, the COMPSs annotations do not interfere with the applications' code, and all applications can be tested sequentially.

\subsubsection{Python Syntax}

The Python syntax in COMPSs is supported through a binding: PyCOMPSs. This Python binding is supported by a Binding-commons layer which focuses on enabling the functionalities of the runtime to other languages (currently, Python and C/C++). It has been designed as an API with a set of defined functions. It is written in C and performs the communication with the runtime through the JNI~\cite{liang1999jni}.

In contrast with the Java syntax, all PyCOMPSs annotations are done inline. The Method Annotations are in the form of Python decorators. Hence, the users can add the \verb|@task| decorator on top of a class or object method to indicate that its invocations will become tasks at execution time. Furthermore, the Parameter Annotations are contained inside the Method Annotation. 

Listing~\ref{listing:sample} shows an example of a task annotation. The first line contains the task annotation in the form of a Python decorator while the rest of the code is a regular python function. The parameter \verb|c| is of type INOUT, and parameters \verb|a|, and \verb|b| are set to the default type IN. The directionality tags are used at execution time to derive the data dependencies between tasks and are applied at an object level, taking into account its references to identify when two tasks access the same object. 

\begin{listing}[!htb]
  \centering
  \footnotesize
  \begin{minted}[xleftmargin=2.7cm, linenos]{python}
@task(c=INOUT)
def multiply(a, b, c):
    c += a * b
  \end{minted}
  \caption{PyCOMPSs Task annotation example.}
  \label{listing:sample}
\end{listing}

A tiny synchronisation API completes the PyCOMPSs syntax. For instance, as shown in Listing~\ref{listing:wordcount}, the \verb|compss_wait_on| waits until all the tasks modifying the \verb|result|'s value are finished and brings the value to the node executing the main program (line 4). Once the value is retrieved, the execution of the main program code is resumed. Given that PyCOMPSs is mostly used in distributed environments, synchronising may imply a data transfer from remote storage or memory space to the node executing the main program.

\begin{listing}[!htb]
  \centering
  \footnotesize
  \begin{minted}[xleftmargin=2.7cm, linenos]{python}
for block in data:
    presult = word_count(block)
    reduce_count(result, presult)
final_result = compss_wait_on(result)
  \end{minted}
  \caption{PyCOMPSs synchronisation API example.}
  \label{listing:wordcount}
\end{listing}

Similarly, the API includes a \verb|compss_open(file_name, mode='r')| to synchronise files, and a \verb|compss_barrier()| to explicitly wait for the completion of all the previous tasks.

%
%
\subsection{Kafka}
\label{subsec:kafka}

\begin{figure}[!htb]
  \centering
  \includegraphics[width=0.9\linewidth]{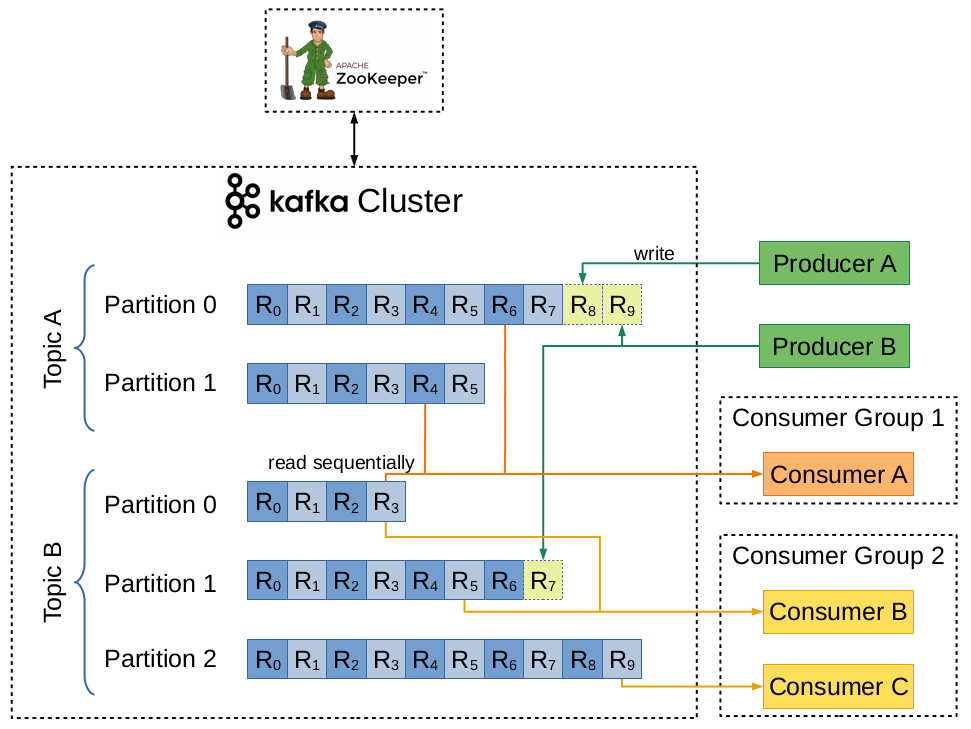}
  \caption{Description of Kafka's basic concepts.}
  \label{fig:kafka_arch}
\end{figure}

Figure~\ref{fig:kafka_arch} illustrates the basic concepts in Kafka and how they relate to each other. \textit{Records} -- each blue box in the figure -- are key-value pairs containing application-level information registered along with its publication time.

Kafka users define several categories or \textit{topics} to which records belong. Kafka relies on ZooKeeper~\cite{zookeeper, web:zookeeper} to store each topic as a partitioned log with an arbitrary number of partitions and maintains a configurable number of partition replicas across the cluster to provide fault tolerance and record-access parallelism. Each partition contains an immutable, publication-time-ordered sequence of records each uniquely identified by a sequential id number known as the \textit{offset} of the record. The example in the figure defines two topics (\textit{Topic A} and \textit{Topic B}) with 2 and 3 partitions, respectively.

Finally, \textit{Producers} and \textit{Consumers} are third-party application components that interact with Kafka to publish and retrieve data. The former add new records to the topics of their choice, while the latter subscribe to one or more topics for receiving records related to them. Consumers can join in \textit{Consumer groups}. Kafka ensures that each record published to a topic is delivered to at least one consumer instance within each subscribing group; thus, multiple processes on remote machines can share the processing of the records of that topic. Although most often delivered exactly once, records might duplicate when one consumer crashes without a clean shutdown and another consumer within the same group takes over its partitions.

Back to the example in the figure, \textit{Producer A} publishes one record to \textit{Topic A}, and \textit{Producer B} publishes two records, one to \textit{Topic A} and one to \textit{Topic B}. \textit{Consumer A}, with a group of its own, processes all the records in \textit{Topic A} and \textit{Topic B}. Since \textit{Consumer B} and \textit{Consumer C} belong to the same consumer group, they share the processing of all the records from \textit{Topic B}. 

Besides the Consumer and Producer API, Kafka also provides the Stream Processor and Connector APIs. The former, usually used in the intermediate steps of the fluent stream processing, allows application components to consume an input stream from one or more topics and produce an output stream to one or more topics. The latter is used for connecting producers and consumers to already existing applications or data systems. For instance, a connector to a database might capture every change to a table. 
    
\section{Architecture}
\label{sec:architecture}

%
%
%

\begin{figure}[!htb]
  \centering
  \includegraphics[width=0.9\linewidth]{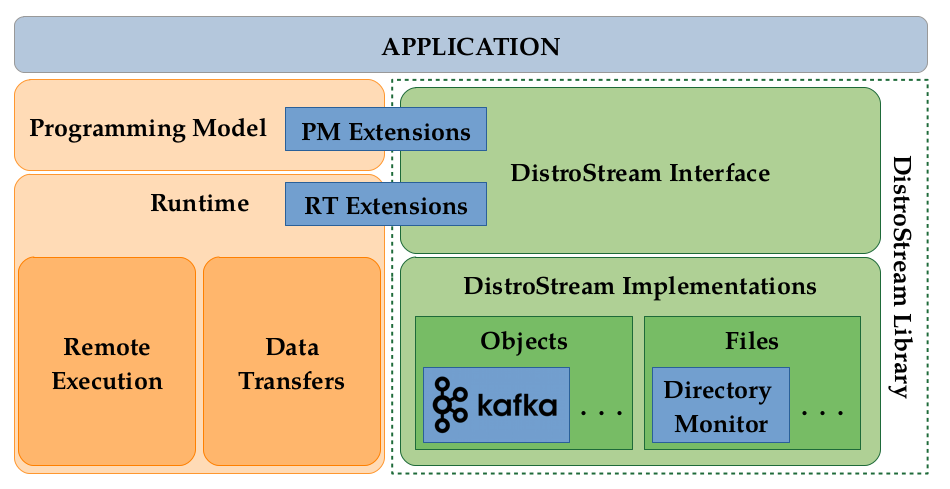}
  \caption{General architecture.}
  \label{fig:general_architecture}
\end{figure}

Figure~\ref{fig:general_architecture} depicts a general overview of the proposed solution. When executing regular task-based workflows, the application written following the programming model interacts with the runtime to spawn the remote execution of tasks and retrieve the desired results. Our proposal includes a representation of a stream (\textit{DistroStream} Interface) that provides applications with homogeneous stream accesses regardless of the stream backend supporting them. Moreover, we extend the programming model and runtime to provide task annotations and scheduling capabilities for streams.

The following subsections discuss the architecture of the proposed solution in a bottom-up approach, starting from the representation of a stream (\textit{DistroStream} API) and its implementations.
Next, we describe the \textit{Distributed Stream Library} and its internal components. Finally, we detail the integration of this library with the programming model (COMPSs) and the necessary extensions of its runtime system. 

%
%
\subsection{Distributed Stream Interface}

The Distributed Stream is a representation of a stream used by applications to publish and receive data values. Its interface provides a common API to guarantee homogeneity on all interactions with streams.

\begin{listing}[!htb]
  \centering
  \footnotesize
  \begin{minted}[xleftmargin=0.7cm, linenos]{java}
// INSTANTIATION
public DistroStream(String alias) throws RegistrationException;

// PUBLISH METHODS  
public abstract void publish(T message) throws BackendException;
public abstract void publish(List<T> messages) throws BackendException;

// POLL METHODS
public abstract List<T> poll() throws BackendException;
public abstract List<T> poll(long timeout) throws BackendException;

// METADATA METHODS
public StreamType getStreamType();
public String getAlias();
public String getId();

// STREAM STATUS
public boolean isClosed();

// CLOSE STREAM
public final void close();
  \end{minted}
\caption{Distributed Stream Interface in Java.}
\label{listing:distro_stream_api}
\end{listing}

As shown in Listing~\ref{listing:distro_stream_api}, the \verb|DistroStream| interface provides a \verb|publish| method for submitting a single message or a list of messages (lines 5 and 6) and a \verb|poll| method to retrieve all the currently available unread messages (lines 9 and 10). Notice that the latter has an optional \verb|timeout| parameter (in milliseconds) to wait until an element becomes available or the specified time expires. Moreover, the streams can be created with an optional \verb|alias| parameter (line 2) to allow different applications to communicate through them. Also, the interface provides other methods to query stream metadata; such as the stream type (line 13), id (line 14), or alias (line 15). Finally, the interface includes methods to check the status of a stream (line 18), and to close it (line 21).

Due to space constraints, Listing~\ref{listing:distro_stream_api} only shows the Java interface, but our prototype also provides the equivalent interface in Python.

%
%
\subsection{Distributed Stream Implementations}

\begin{figure}[!htb]
  \centering
  \includegraphics[width=0.6\linewidth]{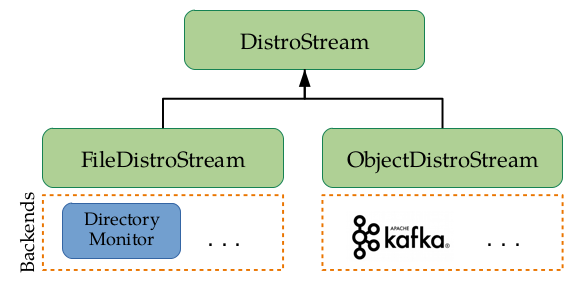}
  \caption{DistroStream class relationship.}
  \label{fig:distro_stream}
\end{figure}

As shown in Figure~\ref{fig:distro_stream}, two different implementations of the \verb|DistroStream| API provide the specific logic to support object and file streams. Object streams are suitable when sharing data within the same language or framework. On the other hand, file streams allow different frameworks and languages to share data. For instance, the files generated by an MPI simulation in C or Fortran can be received through an stream and processed in a Python or Java application.

\subsubsection{Object streams}

\verb|ObjectDistroStream| (ODS) implements the generic \verb|DistroStream| interface to support object streams. Each ODS has an associated \verb|ODSPublisher| and \verb|ODSConsumer| that interact appropriately with the software handling the message transmission (streaming backend). The ODS instantiates them upon the first invocation of a publish or a poll method respectively. This behaviour guarantees that the same object stream has different publisher and consumer instances when accessed from different processes, and that the producer and consumer instances are only registered when required, avoiding unneeded registrations on the streaming backend.

At its current state, the available implementation is backed by Kafka, but the design is prepared to support many backends. Notice that the ODS, the \verb|ODSPublisher|, and the \verb|ODSConsumer| are just abstractions to hide the interaction with the underlying backend. Hence, any other backend (such as an MQTT broker) can be supported without any modification at the workflow level by implementing the functionalities defined in these abstractions. 

Considering the Kafka concepts introduced in Section~\ref{subsec:kafka}, each ODS becomes a Kafka topic named after the stream id. When created, the \verb|ODSPublisher| instantiates a \verb|KafkaProducer| whose \verb|publish| method builds a new \verb|ProducerRecord| and submits it to the corresponding topic via the \verb|KafkaProducer.send| method. If the \verb|publish| invocation sends several messages, the \verb|ODSPublisher| iteratively performs the publishing process for each message so that Kafka registers it as separated records.

Likewise, a new \verb|KafkaConsumer| is instantiated along with an \verb|ODSConsumer|. Then, the \verb|KafkaConsumer| is registered to a consumer group shared by all the consumers of the same application to avoid replicated messages, and subscribed to the topic named after the id of the stream. Hence, the \verb|poll| method retrieves a list of \verb|ConsumerRecords| and deserialises their values. To ensure that records are processed exactly-once, consumers also interact with Kafka's \verb|AdminClient| to delete all the processed records from the database.

\begin{listing}[!htb]
  \centering
  \footnotesize
  \begin{minted}[xleftmargin=0.7cm, linenos]{java}
// PRODUCER
void produce(List<T> objs) {
  // Create stream (alias is not mandatory)
  String alias = "myStream";
  ObjectDistroStream<T> ods = new ObjectDistroStream<>(alias);
  // Metadata getters
  System.out.println("Stream Id: " + ods.getId());
  System.out.println("Stream Alias: " + ods.getAlias());
  System.out.println("Stream Type: " + ods.getStreamType());
  // Publish (single element or list)
  for (T obj : objs) {
    ods.publish(obj);
  }
  ods.publish(objs);
  // Close stream
  ods.close()
}

// CONSUMER
void consume(ObjectDistroStream<T> ods) {
  // Poll current elements (without timeout)
  if (!ods.isClosed()) {
    List<T> newElems = ods.poll();
  }
  // Poll until stream is closed (with timeout)
  while (!ods.isClosed()) {
    List<T> newElems = ods.poll(5)
  }
}
  \end{minted}
\caption{Object Streams (ODS) example in Java.}
\label{listing:streams_java}
\end{listing}

Listing~\ref{listing:streams_java} shows an example using object streams in Java. Notice that the stream creation (line 5) forces all the stream objects to be of the same type \verb|T|. Internally, the stream serialises and deserialises the objects so that the application can publish and poll elements of type \verb|T| directly to/from the stream. As previously explained, the example also shows the usage of the \verb|publish| method for a single element (line 12) or a list of elements (line 14), the \verb|poll| method with the optional \verb|timeout| parameter (lines 23 and 27, respectively), and the common API calls to close the stream (line 16), check its status (line 22), and retrieve metadata information (lines 7 to 9).

Due to space constraints, the example only shows the ODS usage in Java, but our prototype provides an equivalent implementation in Python.

\subsubsection{File streams}

The \verb|FileDistroStream| implementation (FDS) backs the \verb|DistroStream| up to support the streaming of files. Like ODS, its design allows using different backends; however, at its current state, it uses a custom implementation that monitors the creation of files inside a given directory. The \textit{Directory Monitor} backend sends the file locations through the stream and relies on a distributed file system to share the file content. Thus, the monitored directory must be available to every client on the same path.

\begin{listing}[!htb]
  \centering
  \footnotesize
  \begin{minted}[xleftmargin=0.7cm, linenos]{java}
// PRODUCER
void produce(String baseDir, List<String> fileNames) throws IOException {
  // Create stream (alias is not mandatory)
  String alias = "myStream";
  FileDistroStream<T> fds = new FileDistroStream<>(alias, baseDir);
  // Publish files (no need to explicitly call the publish
  // method, the baseDir directory is automatically monitored)
  for (String fileName : fileNames) {
    String filePath = baseDir + fileName;
    try (BufferedWriter writer = new BufferedWriter(new FileWriter(filePath))) {
      writer.write(...);
    }
  }
  // Close stream
  fds.close()
}

// CONSUMER
void consume(FileDistroStream<T> fds) {
  // Poll current elements (without timeout)
  if (!fds.isClosed()) {
    List<String> newFiles = fds.poll();
  }
  // Poll until stream is closed (with timeout)
  while (!fds.isClosed()) {
    List<String> newFiles = fds.poll(5)
  }
}
  \end{minted}
\caption{File Streams (FDS) example in Java.}
\label{listing:streams_python}
\end{listing}

Listing~\ref{listing:streams_python} shows an example using file streams in Java. Notice that the FDS instantiation (line 5 in the listing) requires a base directory to monitor the creation of files and that it optionally accepts an alias argument to retrieve the content of an already existing stream. Also, files are not explicitly published on the stream since the base directory is automatically monitored (lines 8 to 13). Instead, regular methods to write files are used. However, the consumer must explicitly call the \verb|poll| method to retrieve a list of the newly available file paths in the stream (lines 22 and 26). As with ODS, applications can also use the common API calls to close the stream (line 15), check its status (lines 21 and 25), and retrieve metadata information.

Due to space constraints, the example only shows the FDS usage in Java, but our prototype provides an equivalent implementation in Python.

%
%
\subsection{Distributed Stream Library}
\label{subsec:distro_stream_lib}

The Distributed Stream Library (\verb|DistroStreamLib|) handles the stream objects and provides three major components. First, the \verb|DistroStream| API and implementations described in the previous sections.

Second, the library provides the DistroStream Client that must be available for each application process. The client is used to forward any stream metadata request to the DistroStream Server or any stream data access to the suitable stream backend (i.e., \textit{Directory Monitor}, or \textit{Kafka}). To avoid repeated queries to the server, the client stores the retrieved metadata in a cache-like fashion. Either the Server or the backend can invalidate the cached values. 

Third, the library provides the \textit{DistroStream Server} process that is unique for all the applications sharing the stream set. The server maintains a registry of active streams, consumers, and producers with the purpose of coordinating any stream data or metadata access. Among other responsibilities, it is in charge of assigning unique ids to new streams, checking the access permissions of producers and consumers when requesting \verb|publish| and \verb|poll| operations, and notifying all registered consumers when the stream has been completely closed and there are no producers remaining.

\begin{figure}[!htb]
  \centering
  \includegraphics[width=1.0\linewidth]{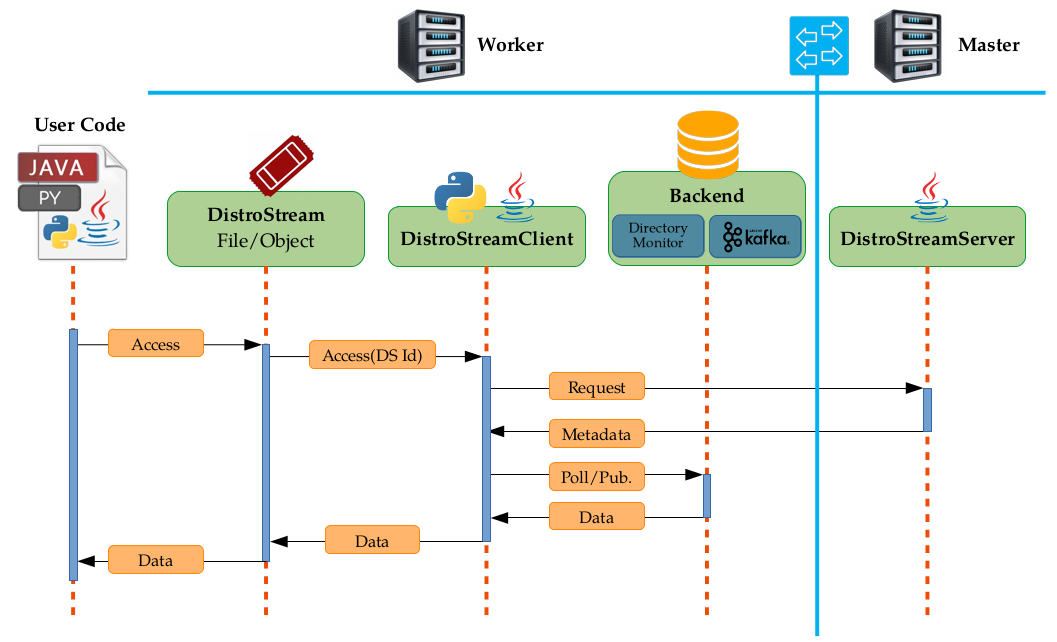}
  \caption{Sequence diagram of the Distributed Stream Library components.}
  \label{fig:distro_stream_lib_seq_diagram}
\end{figure}

Figure~\ref{fig:distro_stream_lib_seq_diagram} contains a sequence diagram that illustrates the interaction of the different Distributed Stream Library components when serving a user petition. The \textit{DistroStream} implementation used by the applications always forwards the requests to the \textit{DistroStream Client} available on the process. The client communicates with the \textit{DistroStream Server} for control purposes, and retrieves from the backend the real data.

%
%
\subsection{Programming Model Extensions}

As already mentioned in Section~\ref{subsec:compss}, the prototype to evaluate Hybrid Workflows is based on the COMPSs workflow manager. At programming-model level, we have extended the COMPSs Parameter Annotation to include a new \textit{STREAM} type. 

As shown in Listing~\ref{listing:stream_param_api}, on the one hand, the users declare producer tasks (methods that write data into a stream) by adding a parameter of type STREAM and direction OUT (lines 3 to 7 in the listing). On the other hand, the users declare consumer tasks (methods that read data from a stream) by adding a parameter of type STREAM and direction IN (lines 9 to 13 in the listing). In the current design, we have not considered INOUT streams because we do not imagine a use case where the same method writes data into its own stream. However, it can be easily extended to support such behaviour when required.

~\newline
~\newline

\begin{listing}[!htb]
  \centering
  \footnotesize
  \begin{minted}[xleftmargin=0.7cm, linenos]{java}
public interface Itf {

    @Method(declaringClass = "Producer")
    Integer sendMessages(
        @Parameter(type = Type.STREAM, direction = Direction.OUT) 
        DistroStream stream
    );
    
    @Method(declaringClass = "Consumer")
    Result receiveMessages(
        @Parameter(type = Type.STREAM, direction = Direction.IN) 
        DistroStream stream
    );
}
  \end{minted}
\caption{Stream parameter annotation example in Java.}
\label{listing:stream_param_api}
\end{listing}

Furthermore, we want to highlight that this new annotation allows integrating streams smoothly with any other previous annotation. For instance, Listing~\ref{listing:stream_hybrid_api} shows a single producer task that uses two parameters: a stream parameter typical of dataflows (lines 5 and 6) and a file parameter typical of task-based workflows (lines 7 and 8).

\begin{listing}[!htb]
  \centering
  \footnotesize
  \begin{minted}[xleftmargin=0.7cm, linenos]{java}
public interface Itf {

    @Method(declaringClass = "Producer")
    Integer sendMessages(
        @Parameter(type = Type.STREAM, direction = Direction.OUT) 
        DistroStream stream,
        @Parameter(type = Type.FILE, direction = Direction.IN) 
        String file
    );    
}
  \end{minted}
\caption{Example combining stream and file parameters in Java.}
\label{listing:stream_hybrid_api}
\end{listing}

%
%
\subsection{Runtime extensions}

\begin{figure}[!htb]
  \centering
  \includegraphics[width=0.95\linewidth]{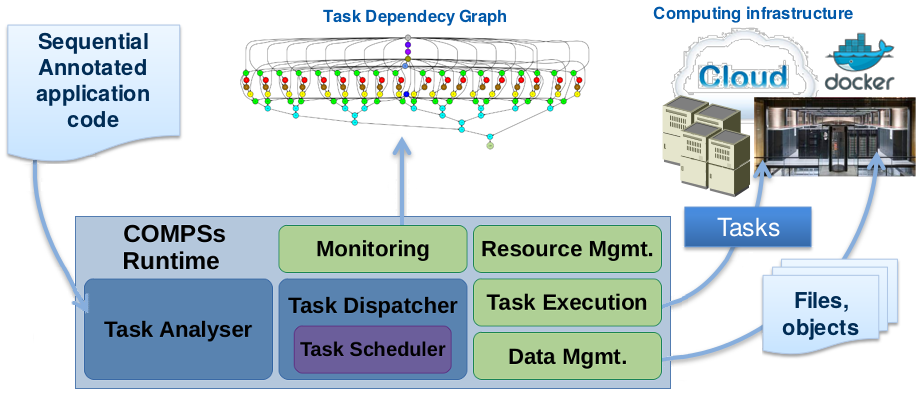}
  \caption{Structure of the internal COMPSs components.}
  \label{fig:compss_internal_components}
\end{figure}

As depicted in Figure~\ref{fig:compss_internal_components}, COMPSs registers the different tasks from the application's main code through the Task Analyser component. Then, it builds a task graph based on the data dependencies and submits it to the \textit{Task Dispatcher}. The Task Dispatcher interacts with the Task Scheduler to schedule the data-free tasks when possible and, eventually, submit them to execution. The execution step includes the job creation, the transfer of the input data, the job transfer to the selected resource, the real task execution on the worker, and the output retrieval from the worker back to the master. If any of these steps fail, COMPSs provides fault-tolerant mechanisms for partial failures. Also, once the task has finished, COMPSs stores the monitoring data of the task, synchronises any data required by the application, releases the data-dependent tasks so that they can be scheduled, and deletes the task.

Therefore, the new Stream annotation has forced modifications in the Task Analyser and Task Scheduler components. More specifically, notice that a stream parameter does not define a traditional data dependency between a producer and a consumer task since both tasks can run at the same time. However, there is some information that must be stored so that the Task Scheduler can correctly handle the available resources and the data locality. In this sense, when using the same stream object, our prototype prioritises producer tasks over consumer tasks to avoid wasting resources when a consumer task is waiting for data to be produced by a non-running producer task. Moreover, the Task Scheduler assumes that the resources that are running (or have run) producer tasks are the data locations for the stream. This information is used to schedule the consumer tasks accordingly and minimise as much as possible the data transfers between nodes. 

\begin{figure}[!htb]
  \centering
  \includegraphics[width=0.9\linewidth]{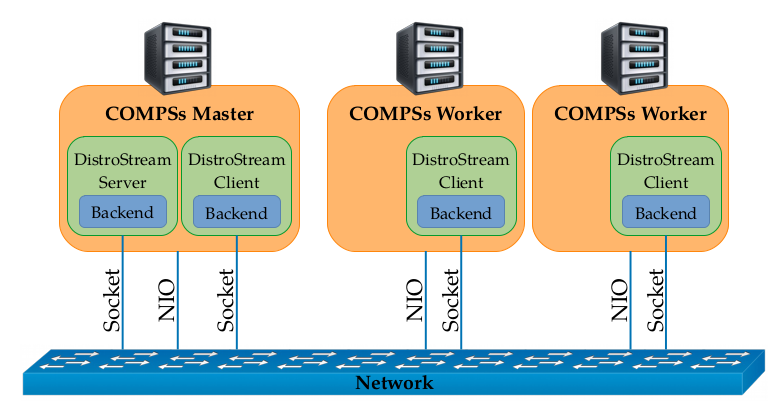}
  \caption{COMPSs and Distributed Stream Library deployment.}
  \label{fig:distro_stream_lib_integration}
\end{figure}

Regarding the components' deployment, as shown in Figure~\ref{fig:distro_stream_lib_integration}, the COMPSs master spawns the \textit{DistroStream Server} and the required backend. Furthermore, it includes a \textit{DistroStream Client} to handle the stream accesses and requests performed on the application's main code. On the other hand, the COMPSs workers only spawn a \textit{DistroStream Client} to handle the stream accesses and requests performed on the tasks. Notice that the COMPSs master-worker communication is done through NIO~\cite{web:nio}, while the \textit{DistroStream Server-Client} communication is done through Sockets.

\section{Use Cases}
\label{sec:use_cases}

%
%
%

Enabling Hybrid task-based workflows and dataflows into a single programming model allows users to define new types of complex workflows. We introduce four patterns that appear in real-world applications so that the users identifying these patterns in their workflows can benefit from the new capabilities and performance improvements of Hybrid Workflows. Next subsections provide in-depth analysis of each use case.

%
%
\subsection{Use case 1: Continuous data generation}
\label{subsec:use_case_1}

One of the main drawbacks of task-based workflows is waiting for task completion to process its results. Often, the output data is generated continuously during the task execution rather than at the end. Hence, enabling data streams allows users to process the data as it is generated.

\begin{listing}[!htb]
  \centering
  \footnotesize
  \begin{minted}[xleftmargin=0.7cm, linenos]{python}
@constraint(computing_units=CORES_SIMULATION)
@task(varargs_type=FILE_OUT)
def simulation(num_files, *args):
    ...
    
@constraint(computing_units=CORES_PROCESS)
@task(input_file=FILE_IN, output_image=FILE_OUT)
def process_sim_file(input_file, output_image):
    ...
    
@constraint(computing_units=CORES_MERGE)
@task(output_gif=FILE_OUT, varargs_type=FILE_IN)
def merge_reduce(output_gif, *args):
    ...
  
def main():
    # Parse arguments
    num_sims, num_files, sim_files, output_files, output_gifs = ...
    # Launch simulations
    for i in range(num_sims):
        simulation(num_files, *sim_files[i])
    # Process generated files
    for i in range(num_sims):
        for j in range(num_files):
            process_sim_file(sim_files[i][j], output_images[i][j])
    # Launch merge phase
    for i in range(num_sims):
        merge_reduce(output_gifs[i], *output_images[i])
    # Synchronize files
    for i in range(num_sims):
        output_gifs[i] = compss_wait_on_file(output_gifs[i])
  \end{minted}
  \caption{Simulations' application in Python without streams.}
  \label{listing:uc1_normal}
\end{listing}

For instance, Listing~\ref{listing:uc1_normal} shows the code of a pure task-based application that launches \verb|num_sims| simulations (line 21). Each simulation produces output files at different time steps of the simulation (i.e., an output file every iteration of the simulation). The results of these simulations are processed separately by the \verb|process_sim_file| task (line 25) and merged to a single GIF per simulation (line 28). The example code also includes the task definitions (lines 1 to 13) and the synchronisation API calls to retrieve the results (line 31).

\begin{figure}[!htb]
  \centering
  \includegraphics[width=0.65\linewidth]{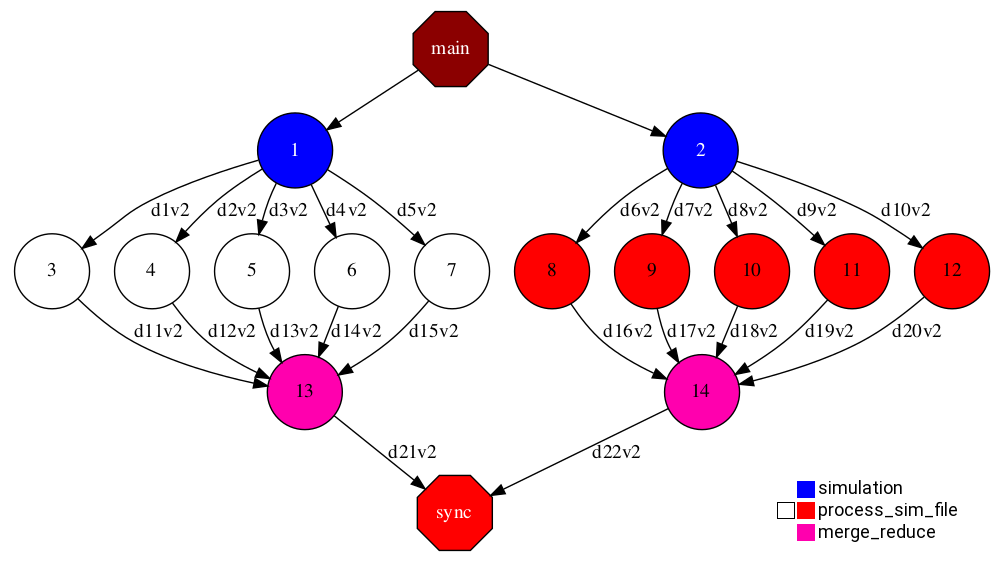}
  \caption{Task graph of the simulation application without streaming.}
  \label{fig:simulation_normal_tg}
\end{figure}

Figure~\ref{fig:simulation_normal_tg} shows the task graph generated by the previous code when running with 2 simulations (\verb|num_sims|) and 5 files per simulation (\verb|num_files|). The \verb|simulation| tasks are shown in blue, the \verb|process_sim_file| in white and red, and the \verb|merge_reduce| in pink. Notice that the simulations and the processing of the files cannot run in parallel since the task-based workflow forces the completion of the simulation tasks to begin any of the processing tasks. 

\begin{listing}[!htb]
  \centering
  \footnotesize
  \begin{minted}[xleftmargin=0.7cm, linenos]{python}
@constraint(computing_units=CORES_SIMULATION)
@task(fds=STREAM_OUT)
def simulation(fds, num_files):
    ...

@constraint(computing_units=CORES_PROCESS)
@task(input_file=FILE_IN, output_image=FILE_OUT)
def process_sim_file(input_file, output_image):
    ...
    
@constraint(computing_units=CORES_MERGE)
@task(output_gif=FILE_OUT, varargs_type=FILE_IN)
def merge_reduce(output_gif, *args):
    ...
    
def main():
    # Parse arguments
    num_sims, num_files, output_images, output_gifs = ...
    # Initialise streams
    input_streams = [None for _ in range(num_sims)]
    for i in range(num_sims):
        input_streams[i] = FileDistroStream(base_dir=stream_dir)
    # Launch simulations
    for i in range(num_sims):
        simulation(input_streams[i], num_files)
    # Process generated files
    for i in range(app_args.num_simulations):
        while not input_streams[i].is_closed():
            # Process new files
            new_files = input_streams[i].poll()
            for input_file in new_files:
                output_image = input_file + ".out"
                output_images[i].append(output_image)
                process_sim_file(input_file, output_image)
    # Launch merge phase
    for i in range(app_args.num_simulations):
        merge_reduce(output_gifs[i], *output_images[i])
    # Synchronise files
    for i in range(app_args.num_simulations):
        output_gifs[i] = compss_wait_on_file(output_gifs[i])
  \end{minted}
  \caption{Simulations' application in Python with streams.}
  \label{listing:uc1_streams}
\end{listing}

On the other hand, Listing~\ref{listing:uc1_streams} shows the code of the same application using streams to retrieve the data from the simulations as it is generated and forwarding it to its processing tasks. The application initialises the streams (lines 20 to 22), launches \verb|num_sims| simulations (line 25), spawns a process task for each received element in each stream (line 34), merges all the output files into a single GIF per simulation (line 37), and synchronises the final results. The \verb|process_sim_file| and \verb|merge_reduce| task definitions are identical to the previous example. Conversely, the \verb|simulation| task definition uses the \verb|STREAM_OUT| annotation to indicate that one of the parameters is a stream where the task is going to publish data. Also, although the simulation, merge, and synchronisation phases are very similar to the pure task-based workflow, the processing phase is completely different (lines 27 to 34). When using streams, the main code needs to check the stream status, retrieve its published elements, and spawn a \verb|process_sim_task| per element. However, the complexity of the code does not increase significantly when adding streams to an existing application. 

\begin{figure}[!htb]
  \centering
  \includegraphics[width=0.65\linewidth]{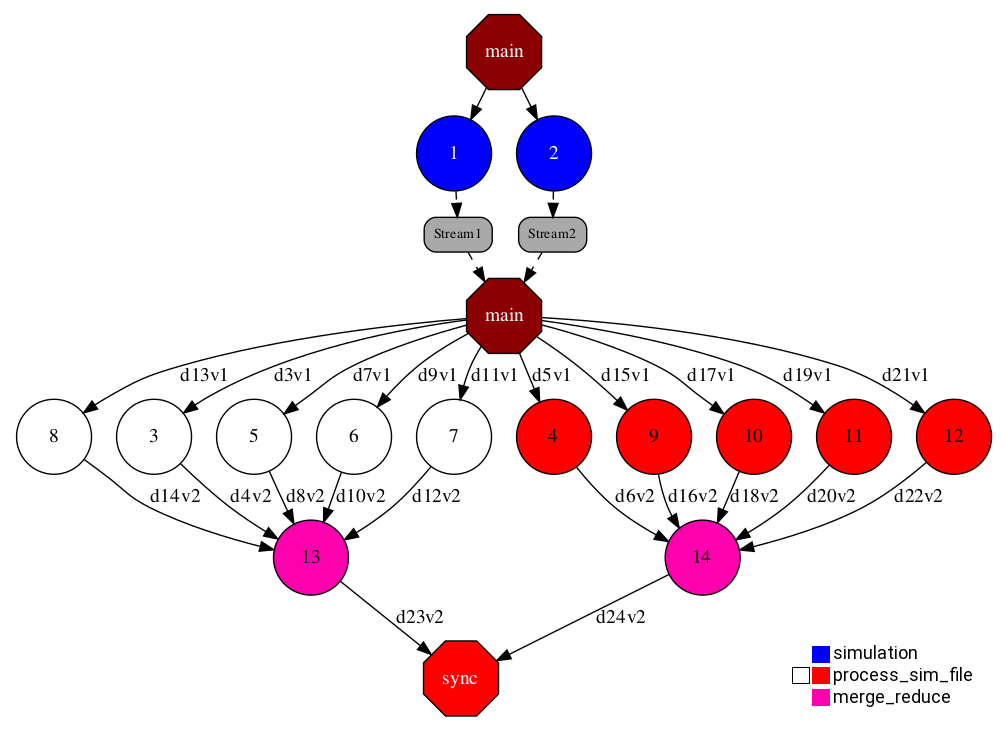}
  \caption{Task graph of the simulation application with streaming.}
  \label{fig:simulation_streams_tg}
\end{figure}

Figure~\ref{fig:simulation_streams_tg} shows the task graph generated by the previous code when running with the same parameters than the pure task-based example (2 simulations and 5 files per simulation). The colour code is also the same than the previous example: the \verb|simulation| tasks are shown in blue, the \verb|process_sim_file| in white and red, and the \verb|merge_reduce| in pink. Notice that streams enable the execution of the processing tasks while the simulations are still running; potentially reducing the total execution time and increasing the resources utilisation (see Section~\ref{subsec:eval_data_cont} for further details).

%
%
\subsection{Use case 2: Asynchronous data exchange}
\label{subsec:use_case_2}

Streams can also be used to communicate data between tasks without waiting for the tasks' completion. This technique can be useful when performing parameter sweep, cross-validation, or running the same algorithm with different initial points.

\begin{figure}[!htb]
  \centering
  \includegraphics[width=0.75\linewidth]{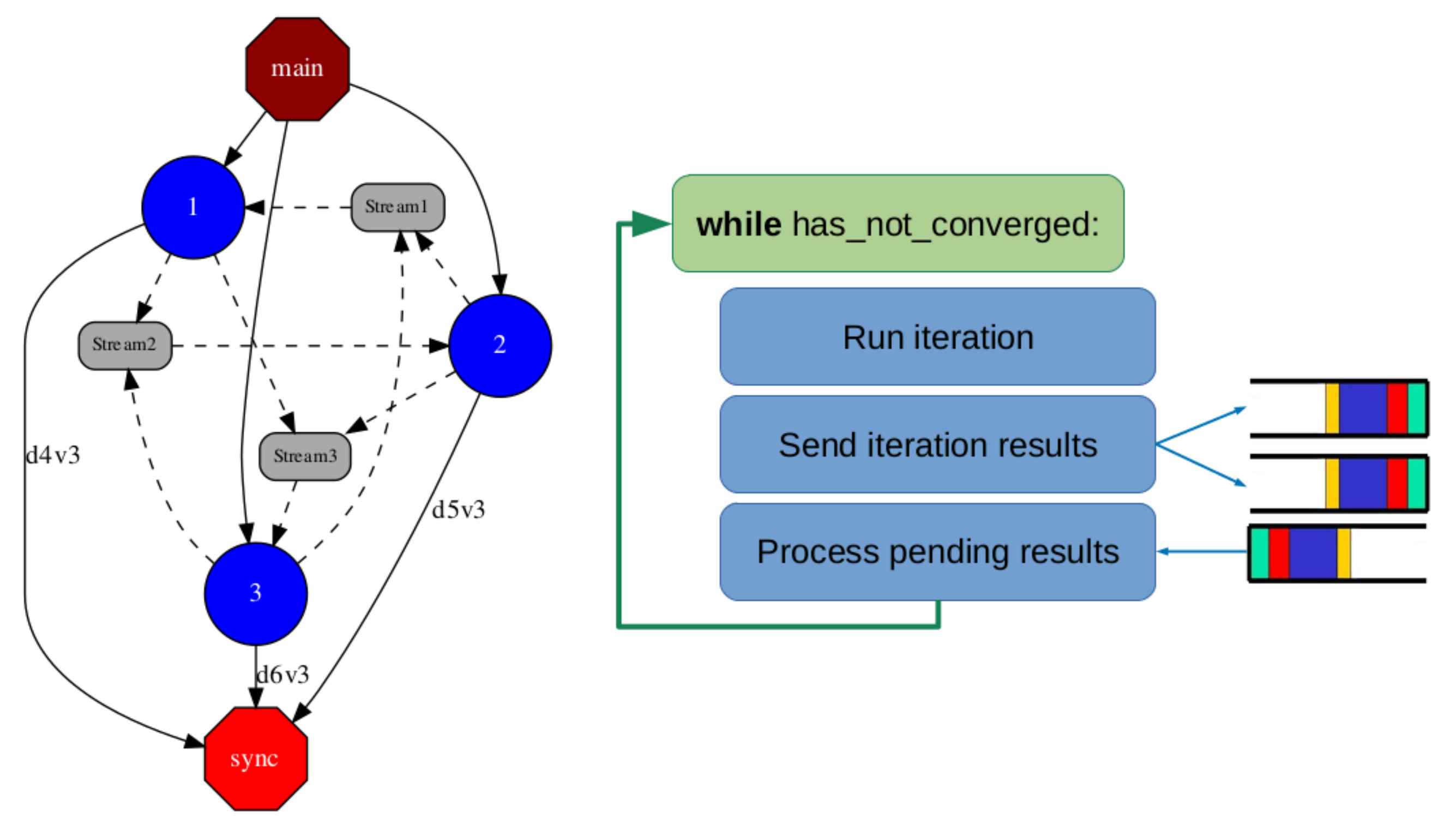}
  \caption{Task graph of the multi-simulations application.}
  \label{fig:multisim_tg}
\end{figure}

For instance, Figure~\ref{fig:multisim_tg} shows three algorithms running simultaneously that exchange control data at the end of every iteration. Notice that the data exchange at the end of each iteration can be done synchronously by stopping all the simulations, or asynchronously by sending the updated results and processing the pending messages in the stream (even though some messages of the actual iteration might be received in the next iteration). Furthermore, each algorithm can run a complete task-based workflow to perform the iteration calculus obtaining a nested task-based workflow inside a pure dataflow.

%
%
\subsection{Use case 3: External streams}

Many applications receive its data continuously from external streams (i.e., IoT sensors) that are not part of the application itself. Moreover, depending on the workload, the stream data can be produced by a single task and consumed by many tasks (one to many), produced by many tasks and consumed by a single task (many to one), or produced by many tasks and consumed by many tasks (many to many). The \textit{Distributed Stream Library} supports all three scenarios transparently, and allows to configure the consumer mode to process the data at least once, at most once, or exactly once when using many consumers.

\begin{figure}[!htb]
  \centering
  \includegraphics[width=0.33\linewidth]{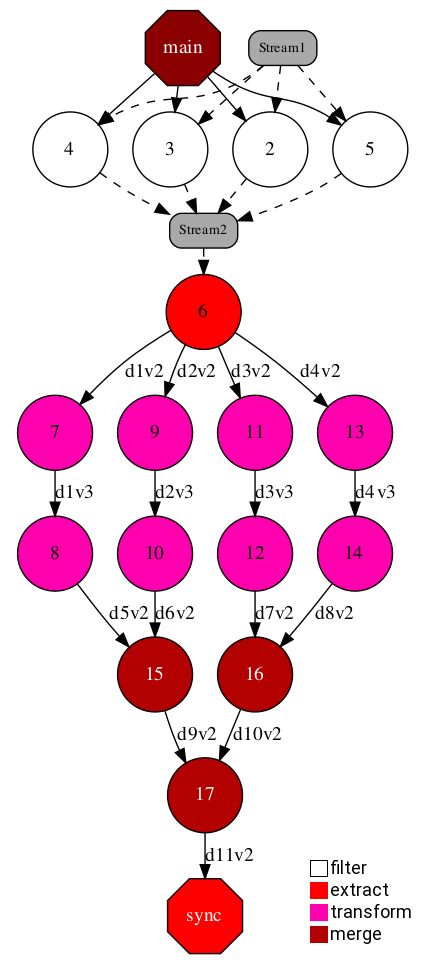}
  \caption{Task graph of the sensor application.}
  \label{fig:sensors_tg}
\end{figure}

Figure~\ref{fig:sensors_tg} shows an external sensor (\textit{Stream 1} in the figure) producing data that is filtered simultaneously by 4 tasks (coloured in white). The relevant data is then extracted from an internal stream (\textit{Stream 2}) by an intermediate task (task 6, coloured in red), and used to run a task-based algorithm. The result is a hybrid task-based workflow and dataflow. Also, the sensor uses a one-to-many stream configured to process the data exactly once, and the \textit{filter} (coloured in white) tasks use a many-to-one stream to publish data to the \textit{extract} task (coloured in red).

%
%
\subsection{Use case 4: Dataflows with nested task-based workflows}

Our proposal also allows to combine task-based workflows and dataflows at different levels; having nested task-based workflows inside a dataflow task or vice-versa. This feature enables the internal parallelisation of tasks, allowing workflows to scale up and down resources depending on the workload. 

\begin{figure}[!htb]
  \centering
  \includegraphics[width=0.6\linewidth]{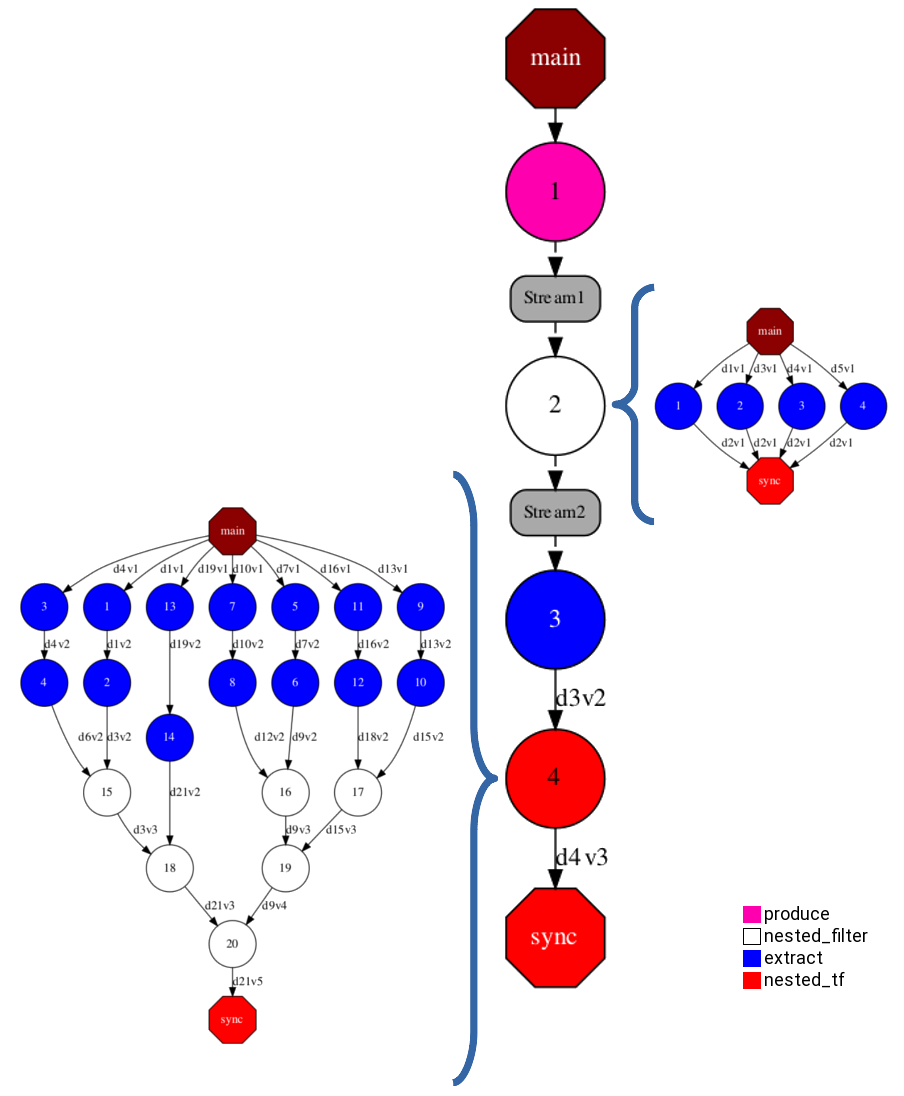}
  \caption{Task graph of the hybrid nested application.}
  \label{fig:hybrid_nested_tg}
\end{figure}

For instance, Figure~\ref{fig:hybrid_nested_tg} shows a dataflow with two nested task-based workflows. The application is similar to the previous use case: the task 1 (coloured in pink) produces the data, task 2 (in white) filters it, task 3 (in blue) extracts and collects the data, and task 4 (in red) runs a big computation.

Notice that, in the previous use case, the application always has 4 filter tasks. However, in this scenario, the filter task has a nested task-based workflow that accumulates the received data into batches and spawns a new filter task per batch. This technique dynamically adapts the resource usage to the amount of data received by the input stream. Likewise, the big computation task also contains a nested task-based workflow. This shows that users can parallelise some computations internally without modifying the original dataflow.
    
\section{Evaluation}
\label{sec:evaluation}

%
%
%

This section evaluates the performance of the new features enabled by our prototype when using data streams against their equivalent implementations using task-based workflows. Furthermore, we analyse the stream writer and reader processes' scalability and load balancing. Finally, we provide an in-depth analysis of the COMPSs runtime performance by comparing the task analysis, task scheduling, and task execution times when using pure task-based workflows or streams. 

%
%
\subsection{Experimental Setup}

The results presented in this section have been obtained using the MareNostrum 4 Supercomputer~\cite{web:mn4} located at the Barcelona Supercomputing Center (BSC). Its current peak performance is 11.15 Petaflops. The supercomputer is composed by 3456 nodes, each of them with two Intel\textregistered Xeon Platinum 8160 (24 cores at 2,1 GHz each). It has 384.75 TB of main memory, 100Gb Intel\textregistered Omni-Path Full-Fat Tree Interconnection, and 14 PB of shared disk storage managed by the Global Parallel File System. 

Regarding the software, we have used DistroStream Library (available at~\cite{github:distro_stream_lib}), COMPSs version 2.5.rc1909 (available at~\cite{github:compss}), and Kafka version 2.3.0 (available at~\cite{github:kafka}). We have also used Java OpenJDK 8 131, Python 2.7.13, GCC 7.2.0, and Boost 1.64.0.

%
%
\subsection{Gain of processing data continuously}
\label{subsec:eval_data_cont}

\begin{figure}[!htb]
\centering
\begin{subfigure}{.95\textwidth}
  \centering
  \includegraphics[width=\linewidth]{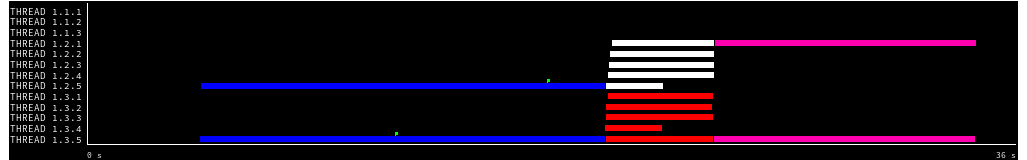}
  \caption{Original COMPSs execution}
\end{subfigure}
\begin{subfigure}{.95\textwidth}
  \centering
  \includegraphics[width=\linewidth]{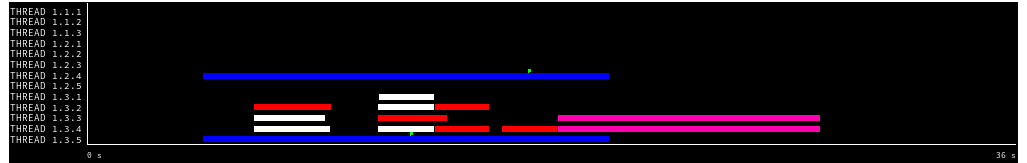}
  \caption{Execution with Hybrid Workflows}
\end{subfigure}
\caption{Paraver traces to illustrate the gain of processing data continuously.}
  \label{fig:simulation_trace}
\end{figure}

As explained in the first use case in Section~\ref{subsec:use_case_1}, one of the significant advantages when using data streams is to process data continuously as it is generated. For that purpose, Figure~\ref{fig:simulation_trace} compares the Paraver~\cite{web:paraver} traces of the original COMPSs execution (pure task-based workflow) and the execution using Hybrid Workflows. Each trace shows the available threads in the vertical axis and the execution time in the horizontal axis - $36 s$ in both traces. Also, each colour represents the execution of a task type; corresponding to the colours shown in the task graphs of the first use case (see Section~\ref{subsec:use_case_1}). The green flags indicate when a simulation has generated all its output files and has closed its associated writing stream. Both implementations are written in Python, and the \textit{Directory Monitor} is set as stream backend.

In contrast to the original COMPSs execution (top), the execution using Hybrid Workflows (bottom) executes the processing tasks (white and red) while the simulations (blue) are still running; significantly reducing the total execution time and increasing the resources utilisation. Moreover, the \verb|merge_reduce| tasks (pink) are able to begin its execution even before the simulation tasks are finished, since all the streams have been closed and the \verb|process_sim_file| tasks have already finished.

In general terms, the gain of the implementation using Hybrid Workflows with respect to the original COMPSs implementation (calculated following Equation~\ref{eq:gain}) is proportional to the number of tasks that can be executed in parallel while the simulation is active. Therefore, we perform an in-depth analysis of the trade-off between the generation and process times. It is worth mentioning that we define the \textit{generation time} as the time elapsed between the generation of two elements of the simulation. Hence, the total duration of the simulation is the generation time multiplied by the number of generated elements. Also, the \textit{process time} is defined as the time to process a single element (that is, the duration of the \verb|process_sim_file| task). 

\begin{equation}
\label{eq:gain}
    Gain = \frac{ExecutionTime_{original} - ExecutionTime_{hybrid}}{ExecutionTime_{original}}
\end{equation}

The experiment uses 2 nodes of 48 cores each. Since the COMPSs master reserves 12 cores, there are two available workers with 36 and 48 cores respectively. The simulation is configured to use 48 cores, leaving 36 available cores while it is active and 84 available cores when it is over. Also, the process tasks are configured to use one single core.

\begin{figure}[!htb]
  \centering
  \includegraphics[width=0.8\linewidth]{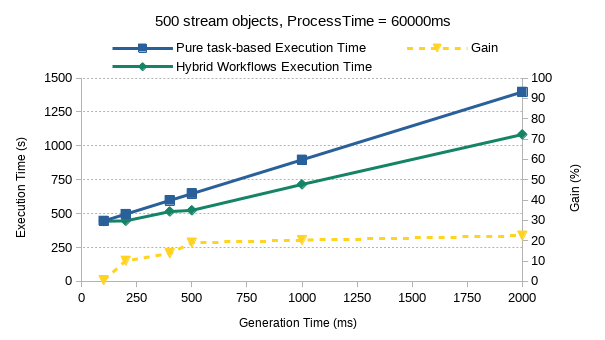}
  \caption{Average execution time and gain of a simulation with increasing generation time.} 
  \label{fig:simulation_generation}
\end{figure}

Figure~\ref{fig:simulation_generation} depicts the average execution time of 5 runs where each simulation generates 500 elements. The process time is fixed to 60,000~ms, while the generation time between stream elements varies from 100~ms to 2,000~ms. For short generation times, almost all the processing tasks are executed when the generation task has already finished, obtaining no gain with respect to the implementation with objects. For instance, when generating elements every 100~ms, the simulation takes 50,000~ms in total ($500 elements \cdot 100 ms/element$). Since the process tasks last 60,000~ms, none of them will have finished before the simulation ends; leading to almost no gain. 

When increasing the generation time, more and more tasks can be executed while the generation is still active; achieving a 19\% gain when generating stream elements every 500~ms. However, the gain is limited because the last generated elements are always processed when the simulation is over. Therefore, increasing the generation time from 500~ms to 2,000~ms only raises the gain from 19\% to 23\%.

\begin{figure}[!htb]
  \centering
  \includegraphics[width=0.8\linewidth]{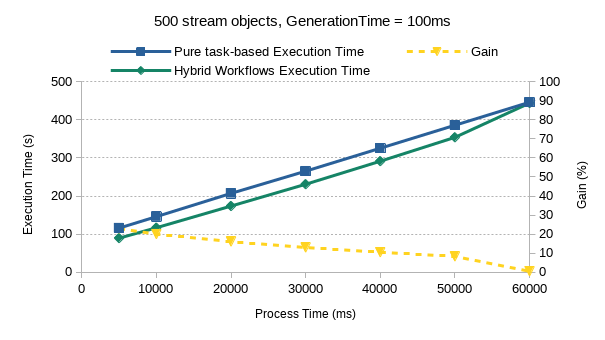}
  \caption{Average execution time and gain of a simulation with increasing process time.} 
  \label{fig:simulation_process}
\end{figure}

On the other hand, Figure~\ref{fig:simulation_process} illustrates the average execution time of 5 runs that generate 500 process tasks with a fixed generation time of 100~ms and a process time varying from 5,000~ms up to 60,000~ms. Notice that the total simulation time is 50,000~ms. When the processing time is short, many tasks can be executed while the generation is still active; achieving a maximum 23\% gain when the processing time is 5,000ms. 

As with the previous case, when the processing time is increased, the number of tasks that can be executed while the generation is active also decreases and, thus, the gain. Also, the gain is almost zero when the processing time is big enough (60,000ms) so that none of the process tasks will have finished before the generation ends. 

\subsection{Gain of removing synchronisations}

Many workflows are composed of several iterative computations running simultaneously until a certain convergence criterion is met. As described in Section~\ref{subsec:use_case_2}, this technique is useful when performing parameter sweep, cross-validation, or running the same algorithm with different initial points.

\begin{figure}[!htb]
  \centering
  \includegraphics[width=0.55\linewidth]{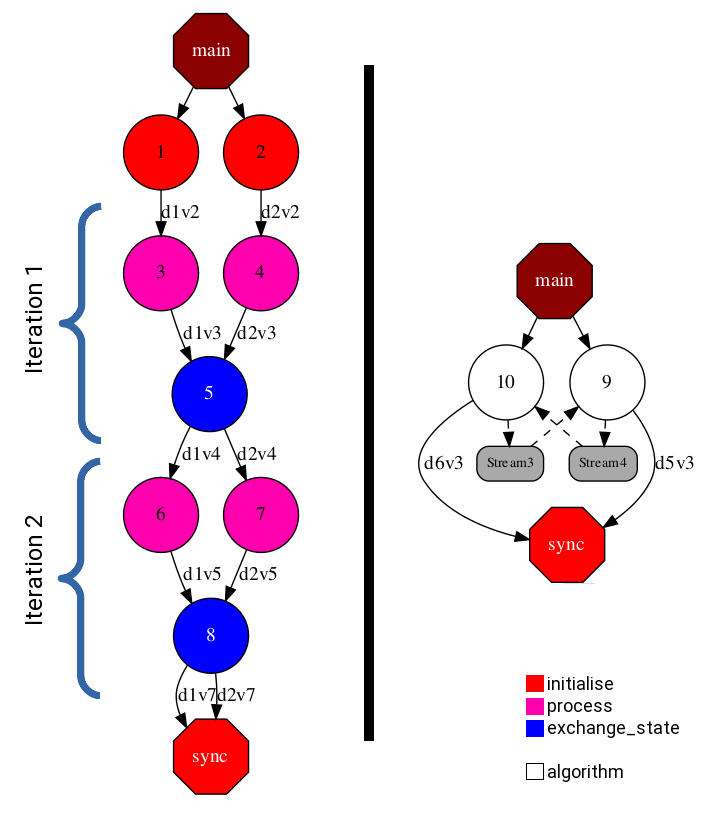}
  \caption{Parallel iterative computations. Pure task-based workflow and Hybrid Workflow shown at left and right, respectively.}
  \label{fig:simsync_task_graphs}
\end{figure}

To this end, each computation requires a phase at the end of each iteration to exchange information with the rest. When using pure task-based workflows, this phase requires to stop all the computations at the end of each iteration, retrieve all the states, create a task to exchange and update all the states, transfer back all the new states, and resume all the computations to the next iteration. The left task graph of Figure~\ref{fig:simsync_task_graphs} shows an example of such workflows with two iterations of two parallel computations. The first two red tasks initialise the state of each computation, the pink tasks perform the computation of each iteration, and the blue tasks retrieve and update the state of each computation.

Conversely, when using Hybrid Workflows, each computation can exchange the information at the end of each iteration asynchronously by writing and reading the states to/from streams. This technique avoids splitting each computation into tasks, stopping and resuming each computation at every iteration, and synchronising all the computations to exchange data. The right task graph of Figure~\ref{fig:simsync_task_graphs} depicts the equivalent Hybrid Workflow of the previous example. Each computation is run in a single task (white) that performs the state initialisation, all the iterations, and all the update phases at the end of each iteration. 

\begin{figure}[!htb]
  \centering
  \includegraphics[width=0.8\linewidth]{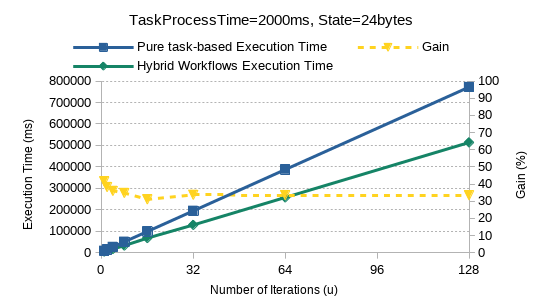}
  \caption{Average execution time and gain of a simulation with an increasing number of iterations.} 
  \label{fig:simsync}
\end{figure}

Using the previous examples, Figure~\ref{fig:simsync} evaluates the performance gain of avoiding the synchronisation and exchange process at the end of each iteration (calculated following Equation~\ref{eq:gain2}). Hence, the benchmark executes the pure task-based workflow (blue) and the Hybrid Workflow (green) versions of the same workflow written in Java and using Kafka as streaming backend. Also, it is composed of two independent computations with a fixed computation per iteration (2,000 ms) and an increasing number of iterations. The results shown are the mean execution times of 5 runs of each configuration.

\begin{equation}
\label{eq:gain2}
    Gain = \frac{ExecutionTime_{pure\: task-based} - ExecutionTime_{hybrid}}{ExecutionTime_{pure\: task-based}}
\end{equation}

~\newline
Notice that the total gain is influenced by three factors: the removal of the synchronisation task at the end of each iteration, the cost of transferring the state between the process and the synchronisation tasks, and, the division of the state's initialisation and process. Although we have reduced the state of each computation to 24 bytes and used a single worker machine to minimise the impact of the transfer overhead, the second and third factors become important when running a small number of iterations (below 32), reaching a maximum gain of 42\% when running a single iteration. For a larger number of iterations (over 32), the removal of the synchronisation becomes the main factor, and the total gain reaches a steady state with a gain around 33\%.

%
%
\subsection{Stream writers and readers scalability and load balance}

Our prototype supports N-M streams, meaning that any stream can have an arbitrary amount of writers and readers. To evaluate the performance and load balance, we have implemented a Java application that uses a single stream and creates N writer tasks and M reader tasks. Although our writer and reader tasks use a single core, we spawn each of them in separated nodes so that the data must be transferred. In more sophisticated use cases, each task could benefit from an intra-node technology (such as OpenMP) to parallelise the processing of the stream data.

\begin{figure}[!htb]
  \centering
  \begin{subfigure}{.5\textwidth}
    \centering
    \includegraphics[width=\linewidth]{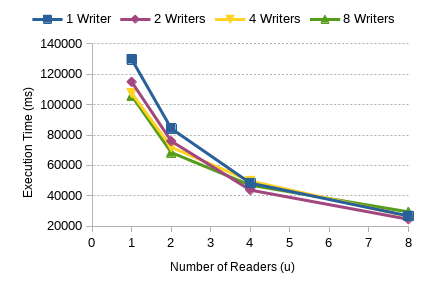}
  \end{subfigure}%
  \begin{subfigure}{.5\textwidth}
    \centering
    \includegraphics[width=\linewidth]{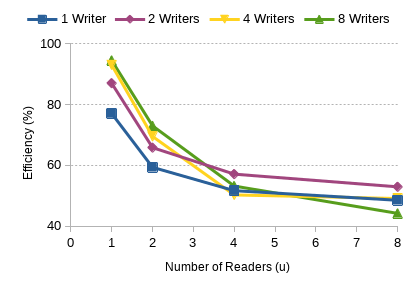}
  \end{subfigure}
  \caption{Average execution time and efficiency (left and right, respectively) with increasing number of readers and different number of writers.}
  \label{fig:scalability_num_wr}
\end{figure}

Figure~\ref{fig:scalability_num_wr} depicts the average execution time (left) and the efficiency (right) of 5 runs with an increasing number of readers. Each series uses a different number of writers, also going from 1 to 8. Also, the writers publish 100 elements in total, the size of the published objects is 24 bytes, and the time to process an element is set to 1,000ms. The efficiency is calculated using the ideal execution time as reference; i.e. the number of elements multiplied by the time to process an element and divided by the number of readers. 

Since the execution time is mainly due to the processing of the elements in the reader tasks, when increasing the number of writers, there are no significant differences. However, for all the cases, increasing the number of readers significantly impacts the execution time, achieving a 4.84 speed-up with 8 readers. Furthermore, the efficiencies using 1 reader are close to the ideal (87\% on average) because the only overheads are the creation of the elements, the task spawning, and the data transfers. However, when increasing the number of readers, the load imbalance significantly affects efficiency; achieving around 50\% efficiency with 8 readers.

\begin{figure}[!htb]
  \centering
  \includegraphics[width=0.9\linewidth]{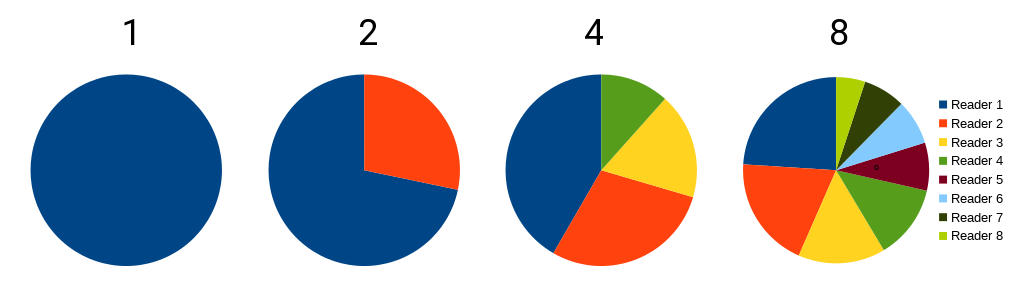}
  \caption{Number of stream elements processed per reader.}
  \label{fig:scalability_elems_per_reader}
\end{figure}

It is worth mentioning that the achieved speed-up is lower than the ideal (8) due to load imbalance. Thus, the elements processed by each reader task are not balanced since elements are assigned to the first process that requests them. Figure~\ref{fig:scalability_elems_per_reader} illustrates an in-depth study of the load imbalance when running 1, 2, 4, or 8 readers. Notice that when running with 2 readers, the first reader gets almost 75\% of the elements while the second one only processes 25\% of the total load. The same pattern is shown when increasing the number of readers; where half of the tasks perform 70\% of the total load. For instance, when running with 4 readers, 2 tasks perform 69\% of the work (34.5\% each), while the rest only performs 31\% of the total load (15.5\% each). Similarly, when running with 8 readers, 4 tasks perform 70\% of the total load (17.5\% each), while the other four only process 30\% (7.5\% each). 

At its current state, the \textit{Distributed Stream Library} does not implement any load balance technique, nor limit the number of elements retrieved by each \verb|poll| call. As future work, since the library already stores the processes registered to each stream, it could implement some policy and send only a subset of the available elements to the requesting process rather than all of them.

%
%
\subsection{Runtime overhead}

To provide a more in-depth analysis of the performance of our prototype, we have compared each step of the task life-cycle when using \verb|ObjectParameter| (OP from now on) or \verb|StreamParameter| (SP from now on) that use \verb|ObjectDistro-| \verb|Streams|. The following figures evaluate the task analysis, task scheduling, and task execution average times of 100 tasks using (a) a single object of increasing size (from 1~MB to 128~MB) or (b) an increasing number of objects (from 1 to 16 objects) of fixed size (8~MB). Both implementations are written in Java and Kafka is used as the stream backend. Regarding the task definition, notice that the OP implementation requires an \verb|ObjectParameter| for each object sent to the task. In contrast, the SP implementation only requires a single \verb|StreamParameter| since all the objects are sent through the stream itself.

\begin{figure}[!htb]
\centering
\begin{subfigure}{.5\textwidth}
  \centering
  \includegraphics[width=\linewidth]{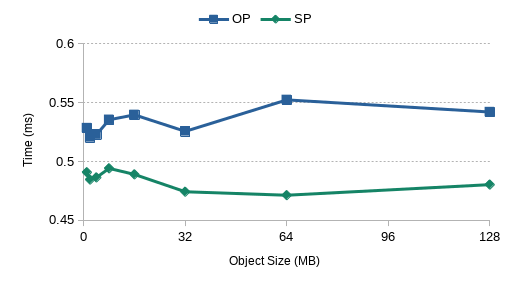}
\end{subfigure}%
\begin{subfigure}{.5\textwidth}
  \centering
  \includegraphics[width=\linewidth]{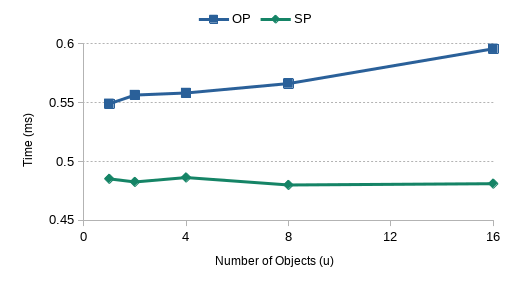}
\end{subfigure}
\caption{Task analysis average time for one single parameter with increasing sizes (left) or increasing number of parameters (right).}
\label{fig:time_bench_ta}
\end{figure}

Figure~\ref{fig:time_bench_ta} compares the task analysis results. The task analysis is the time spent by the runtime to register the task and its parameters into the system. It is worth mentioning that increasing the object's size does not affect the analysis time in either the OP nor the SP implementations. There is, however, a difference around 0.05 ms due to the creation of internal structures to represent object parameters or stream parameters. 

On the other hand, increasing the number of objects directly affects the task analysis time because the runtime needs to register each task parameter individually. For the OP implementation, each object maps to an extra task parameter, and thus, the task analysis time slightly increases when increasing the number of objects. Conversely, for the SP implementation, the stream parameter itself is not modified since we only increase the number of published objects. Hence, the task analysis time remains constant when increasing the number of objects.

\begin{figure}[!htb]
\centering
\begin{subfigure}{.5\textwidth}
  \centering
  \includegraphics[width=\linewidth]{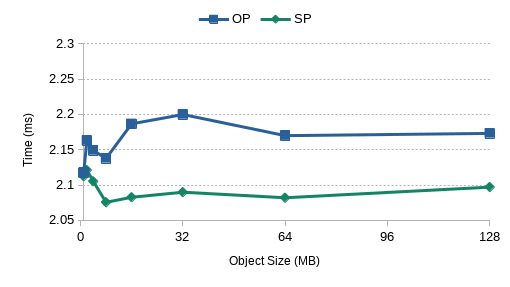}
\end{subfigure}%
\begin{subfigure}{.5\textwidth}
  \centering
  \includegraphics[width=\linewidth]{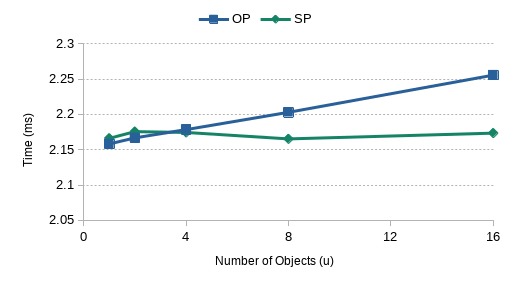}
\end{subfigure}
\caption{Task scheduling average time for one single parameter with increasing sizes (left) or increasing number of parameters (right).}
\label{fig:time_bench_td}
\end{figure}

Figure~\ref{fig:time_bench_td} compares the task scheduling results. On the one hand, the scheduling time for both implementations varies from 2.05~ms to 2.20~ms but does not show any clear tendency to increase when increasing the object's size. On the other hand, when increasing the number of objects, the scheduling time increases for the OP implementation and remains constant for the SP implementation. This behaviour is due to the fact that the default COMPSs scheduler implements data locality and, thus, the scheduling time is proportional to the number of parameters. Similarly to the previous case, increasing the number of objects increases the number of task parameters for the OP implementation (increasing its scheduling time), but keeps a single parameter for the SP implementation (maintaining its scheduling time).

\begin{figure}[!htb]
\centering
\begin{subfigure}{.5\textwidth}
  \centering
  \includegraphics[width=\linewidth]{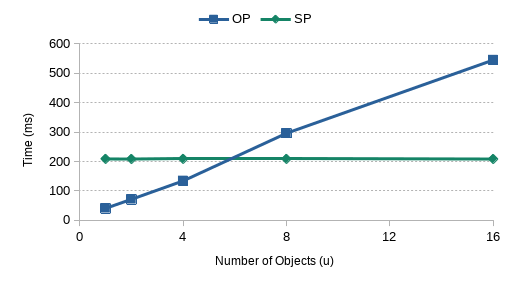}
\end{subfigure}%
\begin{subfigure}{.5\textwidth}
  \centering
  \includegraphics[width=\linewidth]{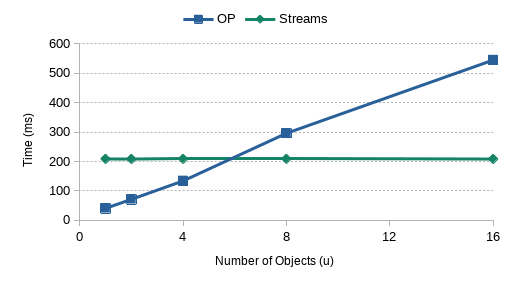}
\end{subfigure}
\caption{Task execution average time for one single parameter with increasing sizes (left) or increasing number of parameters (right).}
\label{fig:time_bench_texec}
\end{figure}

Figure~\ref{fig:time_bench_texec} compares the task execution results. The task execution time covers the transfer of all the task parameters and the task execution itself. Regarding the SP implementation, the time remains constant around 208~ms regardless of the object's size and the number of objects because the measurement only considers the transfer of the stream object itself and the execution time of the \verb|poll| method. It is worth mentioning that the actual transfers of the objects are done by Kafka when invoking the \verb|publish| method on the main code, and thus, they are executed in parallel while COMPSs spawns the task in the worker machine.

Conversely, the execution time for the OP implementation increases with both the object's size and the number of objects, since the serialisation and transfer times also increase. However, the task execution does not need to fetch the objects (the \verb|poll| method) since all of them have already been transferred. This trade-off can be observed in the figure, where the OP implementation performs better than the SP implementation when using task parameters smaller than 48~MB and performs worse for bigger cases. Notice that only the total objects' size is relevant since the same behaviour is shown when using a single 48~MB object or 6 objects of 8~MB each. 

\begin{figure}[!htb]
  \centering
  \includegraphics[width=0.7\linewidth]{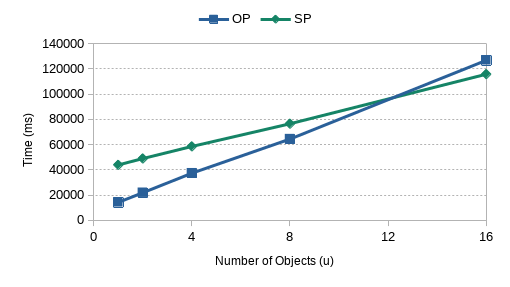}
  \caption{Total execution time with increasing number of parameters.}
  \label{fig:time_bench_ttotal_numobjs}
\end{figure}

Since the real object transfers when using SP are executed during the \verb|publish| method and cannot be observed measuring the task execution time, we have also measured the total execution time of the benchmark for both implementations. Figure~\ref{fig:time_bench_ttotal_numobjs} shows the total execution time with an increasing number of objects of 8~MB. In contrast to the previous plot, both implementations have an increasing execution time proportional to the objects' size. Also, the SP implementation only outperforms the OP implementation when using more than 12 objects.

To conclude, since there are no major differences regarding the task analysis time nor the task scheduling time, we can safely assume that the use of streams instead of regular objects is recommended when the total size of the task parameters exceeds 48~MB and there are more than 12 objects published to the stream. 
    
\section{Conclusion and Future Work}
\label{sec:conclusion}

%
%
%


This paper demonstrates that task-based workflows and dataflows can be integrated into a single programming model to better-cover the needs of the new Data Science workflows. Using Hybrid Workflows, developers can build complex pipelines with different approaches at many levels using a single framework. 

The proposed solution relies on the \textit{DistroStream} concept: a generic API used by applications to handle stream accesses homogeneously regardless of the software backing it. Two implementations provide the specific logic to support object and file streams. The first one, \textit{ObjectDistroStream}, is built on top of Kafka to enable object streams. The second one, \textit{FileDistroStream}, monitors the creation of files inside a directory, sends the file locations through the stream, and relies on a distributed file system to share the file content.

The \textit{DistroStream} API and both implementations are part of the \textit{DistroStreamLib}, which also provides the \textit{DistroStream Client} and the \textit{DistroStream Server}. While the client acts as a broker on behalf of the application and interacts with the corresponding backend, the server manages the streams' metadata and coordinates the accesses.

By integrating the \textit{DistroStreamLib} into a task-based Workflow Manager, its programming model can easily support Hybrid Workflows. The described prototype extends COMPSs to enable tasks with continuous input and output data by providing a new annotation for stream parameters. Implementing the handling of such stream-type values lead to some modifications on the Task Analyser and Task Scheduler components of the runtime. Using the \textit{DistroStreamLib} also implied changes at deployment time since its components need to be spawned along with COMPSs. On the one hand, the COMPSs master hosts the \textit{DistroStream Server}, the required stream backend, and a \textit{DistroStream Client} to handle stream accesses on the application's main code. On the other hand, each COMPSs worker contains a \textit{DistroStream Client} that performs the stream accesses on tasks. Although the described prototype only builds on COMPSs, it can be used as an implementation reference for any other existing task-based framework.

This paper also presents four use cases illustrating the new capabilities that the users may identify in their workflows to benefit from the use of Hybrid Workflows. On the one hand, streams can be internal or external to the application and can be used to communicate continuous data or control data. On the other hand, streams can be accessed inside the main code, native tasks (i.e., Java or Python), or non-native tasks (i.e., MPI, binaries, and nested COMPSs workflows). Furthermore, the \textit{Distributed Stream Library} supports the one to many, many to one, and many to many scenarios transparently, and allows to configure the consumer mode to process the data at least once, at most once, or exactly once when using many consumers.

The evaluation demonstrates the benefit of processing data continuously as it is generated; achieving a 23\% gain with the right generation and process times and resources. 
Also, using streams as control mechanism enabled the removal of synchronisation points when running several parallel algorithms, leading to a 33\% gain when running more than 32 iterations.
Finally, an in-depth analysis of the runtime's performance shows that there are no major differences regarding the task analysis time nor the task scheduling time when using streams or object tasks, and that the use of streams is recommended when the total size of the task parameters exceeds 48 MB using more than 12 objects.


Although the solution is fully functional, some improvements can be made. Regarding the \textit{DistroStream} implementations, we plan to extend the \textit{FileDistroStream} to support shared disks with different mount-points. On the other hand, we will add new \textit{ObjectDistroStream}'s backend implementations (apart from Kafka), so that users can choose between them without changing the application's code. Envisaging that a single \textit{DistroStream Server} could become a bottleneck when managing several applications involving a large number of cores, we consider replacing the client-server architecture by a peer-to-peer approach. Finally, by highlighting the benefits from hybrid flows, we expect to attract real-world applications to elevate our evaluation to more complex use cases.

    \section*{Acknowledgements}

This work has been supported by the Spanish Government (contracts SEV2015-0493 and TIN2015-65316-P), by Generalitat de Catalunya (contract 2014-SGR-1051), and by the European Commission through the Horizon 2020 Research and Innovation program under contract 730929 (MF2C project). Cristian Ramon-Cortes predoctoral contract is financed by the Spanish Government under the contract BES-2016-076791.

    \section*{References}
    \bibliographystyle{elsarticle-num}
    \bibliography{streaming}

\end{document}